\documentclass[letter,twocolumn]{jpsj3} 
\usepackage{bm,amssymb,amsmath}	
\usepackage{graphicx}   
\usepackage{txfonts}
\newcommand{\simg}{\stackrel{>}{_\sim}}
\newcommand{\siml}{\stackrel{<}{_\sim}}

\title{
FFLO Excitonic State in the Three-Chain Hubbard Model for Ta$_2$NiSe$_5$
}
\author{Takemi {\sc Yamada}$^1$\thanks{E-mail address: takemi@phys.sc.niigata-u.ac.jp}, Kaoru {\sc Domon}$^2$, and Yoshiaki {\sc \=Ono}$^2$}
\inst{
$^1$CFIL, Niigata University, Ikarashi, Nishi-ku, Niigata, 950-2181, Japan \\
$^2$Department of Physics, Niigata University, Ikarashi, Nishi-ku, Niigata, 950-2181, Japan}

\abst{
The three-chain Hubbard model for Ta$_2$NiSe$_5$, known as a candidate material for an excitonic insulator, is investigated over the wide range of the energy gap $D$  between the twofold degenerate conduction bands and the nondegenerate valence band including both semiconducting ($D>0$) and semimetallic ($D<0$) cases. In the semimetallic case, the difference in the band degeneracy inevitably causes the imbalance of each Fermi wavenumber, resulting in a remarkable excitonic state characterized by the condensation of excitons with finite center-of-mass momentum $q$, the so-called Fulde-Ferrell-Larkin-Ovchinnikov (FFLO) excitonic state. With decreasing $D$ corresponding to increasing pressure, the obtained excitonic phase diagram shows a crossover from BEC ($D\simg 0$) to BCS ($D\siml 0$) regime, and then shows a distinct phase transition at a certain critical value $D_c(<0)$ from the uniform ($q=0$) to the FFLO ($q\ne 0$) excitonic state, as expected to be observed in Ta$_2$NiSe$_5$ under high pressure. 
}

\begin{document}
\maketitle

Recently, Ta$_2$NiSe$_5$ has attracted much attention as a strong candidate for the excitonic insulator (EI) which is characterized by the condensation of excitons and has been argued since about half a century ago\cite{Knox.1963,PR.158.462,RMP.40.755}. Its resistivity indicates that it is a narrow-gap semiconductor with a quasi-one-dimensional (1D) structure, where Ni and Ta atoms are arranged in 1D chains\cite{Inorg.Chem.24.3611,JLCM.116.51}. 
A structural transition from the orthorhombic to monoclinic phase occurs at $T_c$=328 K\cite{JLCM.116.51}, below which the magnetic susceptibility shows a gradual drop, and flattening of the valence band top has been observed in the ARPES experiments\cite{PRL.103.026402,JSNM.25.1231}. 
Several theoretical studies\cite{PRB.87.035121,PRB.93.041105,PRB.90.155116} have revealed that the transition can be interpreted as excitonic condensation from a normal semiconductor to the excitonic insulator from a mean-field analysis for the three-chain Hubbard model with electron-phonon coupling\cite{PRB.87.035121,PRB.93.041105} and from a variational cluster approximation for the extended Falicov-Kimball model\cite{PRB.90.155116}.

\begin{figure}[b]
\begin{center}
\includegraphics[width=7.0cm]{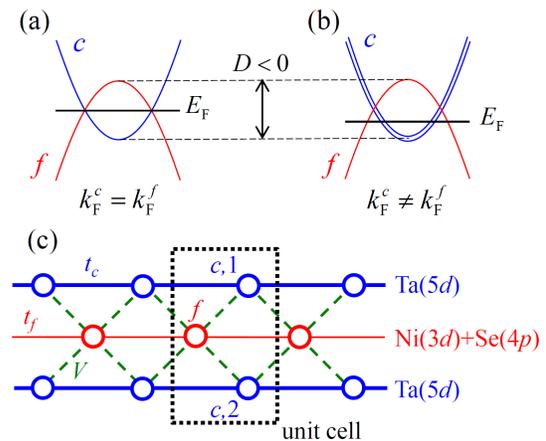}
\caption{(Color online) 
Semimetallic band structures with a negative energy gap $D$ for total electron number $n=2$ in the cases that both conduction ($c$) and valence ($f$) bands are nondegenerate where $k_{\rm F}^{c}=k_{\rm F}^{f}$ (a), and the $c$ band is twofold degenerate while the $f$ band is nondegenerate where $k_{\rm F}^{c}\neq k_{\rm F}^{f}$ (b), as expected to be realized in Ta$_2$NiSe$_5$ under high pressure. 
(c) Schematic representation of the three-chain Hubbard model for Ta$_2$NiSe$_5$\cite{PRB.87.035121}.
}
\label{Fig1}
\end{center}
\end{figure}

Usually, excitonic condensations have been discussed in a narrow-gap semiconductor or a semimetal with slight band overlapping with nondegenerate conduction and valence bands for simplicity\cite{Knox.1963,PR.158.462,RMP.40.755}, where each Fermi wavevector in the semimetallic case coincides to each other as shown in Fig. \ref{Fig1}(a). However, the band structure calculation\cite{PRB.87.035121} revealed that twofold degenerate conduction bands exist in Ta$_2$NiSe$_5$, originating from two Ta 5$d$ orbitals, while the nondegenerate valence band originates from hybridized Ni 3$d$ and Se 4$p$ orbitals as shown in 
Fig. \ref{Fig1}(c). The difference in the band degeneracy inevitably causes the imbalance of each Fermi wavenumber in the semimetallic case, as shown in Fig. \ref{Fig1}(b), where one can expect that the condensation of excitons with finite center-of-mass momentum $q$ takes place, analogous to that of Cooper pairs in Fulde-Ferrell-Larkin-Ovchinnikov (FFLO) superconductivity under an external magnetic field, where the Zeeman splitting causes the imbalance of the Fermi wavenumber for each spin. In fact, several authors have recently discussed the possibility of the FFLO excitonic state in the electron-hole bilayer systems with density imbalance\cite{PRB.75.113301,JPSJ.79.033001,PRB.81.115329}. 
The purpose of this letter is to clarify what kind of excitonic phase (EP) exists in the semimetallic case of the three-chain Hubbard model for Ta$_2$NiSe$_5$, which has not been discussed in the previous theoretical studies\cite{PRB.87.035121,PRB.93.041105} but might be realized in experiments under high pressure\cite{pressure}.

The three-chain Hubbard model for Ta$_2$NiSe$_5$\cite{PRB.87.035121} consists of the twofold degenerate conduction ($c$) bands from Ta $5d$ orbitals and the nondegenerate valence ($f$) band from hybridized Ni $3d$ and Se $4p$ orbitals as schematically shown in Figs. \ref{Fig1}(b) and \ref{Fig1}(c). Its Hamiltonian is explicitly given by $H=H_0+H'$ with 
\begin{align}
H_0&=\sum_{k\sigma}\sum_{\alpha=1,2}\epsilon_{k}^{c}c^{\dagger}_{k\alpha\sigma}c_{k\alpha\sigma}
+\sum_{k\sigma}\epsilon_{k}^{f}f^{\dagger}_{k\sigma}f_{k\sigma}, \label{eq:H0}\\
H'&=V\sum_{i\alpha}\sum_{\sigma\sigma^\prime}\left(
c^{\dagger}_{i-1\alpha\sigma}c_{i-1\alpha\sigma}
+c^{\dagger}_{i\alpha\sigma}c_{i\alpha\sigma}\right)
f^{\dagger}_{i\sigma^\prime}f_{i\sigma^\prime},
\label{eq:H1}
\end{align}
where $c_{k\alpha\sigma}(c_{i\alpha\sigma})$ and $f_{k\sigma}(f_{i\sigma})$ are the annihilation operators for $c$ and $f$ electrons with wavenumber $k$ (site $i$), spin $\sigma=\uparrow,\downarrow$ and chain degrees of freedom for the $c$ electron $\alpha=1,2$. 
The noninteracting $c(f)$ band dispersion is given by 
\[
\epsilon_{k}^{c(f)}=2t_{c(f)}\left({\rm cos}k-1\right)+(-) D/2, 
\]
where $t_{c}$ and $t_{f}$ are the $c$ and $f$ hopping parameters and set to $t_{c}=-0.8$ eV and $t_{f}=0.4$ eV, respectively, which have been determined in Ref.\cite{PRB.87.035121} so as to fit the energy band from the first-principles calculation for Ta$_2$NiSe$_5$. $D$ is the energy gap between the $c$ and $f$ bands at $k=0$, describing both semiconducting ($D>0$) and semimetallic ($D<0$) cases. As $D$ is considered to be a decreasing function of pressure, we vary $D$ as a parameter instead of fixing $D$ to 0.2 eV as in Ref.\cite{PRB.87.035121} so as to reproduce the first-principles energy band at ambient pressure. 
In Eq. (\ref{eq:H1}), we consider the intersite $c$-$f$ Coulomb interaction $V$ which is crucial for the excitonic order as shown below, while we neglect the on-site Coulomb interaction, which can be effectively included in $D$ and/or the chemical potential $\mu$ within the mean-field approximation by excluding the magnetic and density-wave-type orders\cite{PRB.87.035121}. 

Now, we discuss the excitonic order within the mean-field approximation in which $H'$ in Eq. (\ref{eq:H1}) is replaced by 
\begin{equation}
H'_{\rm MF}=
\sum_{kq\sigma}\sum_{\alpha=1,2}\left(\Delta(k,q)c_{k\alpha\sigma}^{\dagger}f_{k+q\sigma}+{\rm H.c.}\right)+{\rm const}.
\nonumber
\end{equation}
Here the excitonic order parameter 
$\Delta(k,q)=-\frac{V}{N}\sum_{k^\prime}(1+e^{i(k-k^\prime)})\langle f^{\dagger}_{k^\prime+q\sigma}c_{k^\prime\alpha\sigma}\rangle$ 
becomes finite when the condensation of excitonic $c$-$f$ pairs with center-of-mass momentum $q$ takes place and is assumed to be independent of $\sigma$ and $\alpha$ for simplicity. Diagonalizing $H_{\rm MF}=H_0+H'_{\rm MF}$ to yield the mean-field band dispersion
\begin{equation}
E_{k,\pm}^{\rm MF}=\epsilon_{+}(k,q)\pm\sqrt{\epsilon_{-}^{2}(k,q)+2|\Delta(k,q)|^{2}}
\label{eq:EMF}
\end{equation}
with $\epsilon_{\pm}(k,q)=(\epsilon_{k}^{c}\pm\epsilon_{k+q}^{f})/2$, we obtain the gap equation to determine $\Delta(k,q)$ as
\begin{align}
&\Delta(k,q)=\frac{V}{N}\sum_{k^\prime}(1+e^{i(k-k^\prime)})\Delta(k^\prime,q)g(k^\prime,q)
\label{eq:OP}
\end{align}
with 
$g(k,q)=\frac12(f(E^{\rm MF}_{k,-})-f(E^{\rm MF}_{k,+}))/\sqrt{\epsilon_{-}^{2}(k,q)+2|\Delta(k,q)|^{2}}$, 
where $f(\epsilon)=1/(e^{(\epsilon-\mu)/k_B T}+1)$. 
In Eq. (\ref{eq:OP}), $\Delta(k,q)$ can be rewritten as 
\begin{equation}
\Delta(k,q)=\Delta_{q}^{(0)}+\Delta_{q}^{(1)}e^{ik}
  =\Delta_{q}(1+e^{ik}e^{-i\phi_{q}}),
\label{eq:Delta}
\end{equation}
where $\Delta_{q}$ is the magnitude of the order parameter and $\phi_{q}$ is the relative phase between the nearest-neighbor $c$-$f$ pair with the $c$-site to the right of the $f$-site $\Delta_{q}^{(0)}$ and that to the left $\Delta_{q}^{(1)}$. Substituting Eq. (\ref{eq:Delta}) into Eq. (\ref{eq:OP}), we obtain the following self-consistent equations to determine $\Delta_{q}$ and $\phi_{q}$ :
\begin{align}
&\chi^{(0)}(q)+|\chi^{(1)}(q)|=1/V,
\label{eq:sce1}\\
&\tan{\phi_{q}}={\rm Im}~\chi^{(1)}(q)/{\rm Re}~\chi^{(1)}(q),
\label{eq:sce2}
\end{align}
where 
$\chi^{(n)}(q)=\frac{1}{N}\sum_{k}e^{ikn}g(k,q)$. 
When we set $q=\phi_{q}=0$ in Eqs. (\ref{eq:sce1}) and (\ref{eq:sce2}), the solution coincides with that in Ref.\cite{PRB.87.035121}, where the semimetallic case ($D<0$) responsible for the finite $q$ ($\phi_{q}$) solution is not considered. 

Generally, Eqs. (\ref{eq:sce1}) and (\ref{eq:sce2}) yield self-consistent solutions of $\Delta_{q}$ and $\phi_{q}$ for various values of $q$. Therefore, we determine the most stable solution by minimizing the free energy 
\begin{align}
&\delta F_{q}(n,T,\Delta_{q},\phi_{q})=F_{q}^{\rm MF}(n,T,\Delta_{q},\phi_{q})-F_{0}(n,T)
\nonumber \\
&=-\frac{T}{N}\sum_{ks\sigma}{\rm ln}\left(
\frac{1+e^{-(E_{ks}^{\rm MF}-\mu)/k_B T}}{1+e^{-(E_{ks}^{0}-\mu_0)/k_B T}}
\right)
+(\mu-\mu_0)n+\frac{8|\Delta_q|^2}{V}
\nonumber
\end{align}
w.r.t. the wavenumber $q$, where $F_0$, $E_{ks}^{0}$, and $\mu_0$ are the free energy, the energy band, and the chemical potential for the normal state with $\Delta_q=\phi_{q}=0$, respectively, $s$ is the band index, and $\mu$ and $\mu_0$ are determined so as to fix the number of electrons per unit cell to $n=n^{c}+n^{f}$. Note that the self-consistent Eqs. (\ref{eq:sce1}) and (\ref{eq:sce2}) can be reproduced by the stationary conditions $\partial \delta F_{q}/\partial \Delta_{q}=0$ and $\partial \delta F_{q}/\partial \phi_q=0$ for a given $q$. 
In the present study, we set $n=2$ and $V=0.4$ eV and vary $T$ and $D$ as parameters. Here and hereafter, the energy is measured in units of eV.

\begin{figure}[t]
\begin{center}
\includegraphics[width=6.5cm]{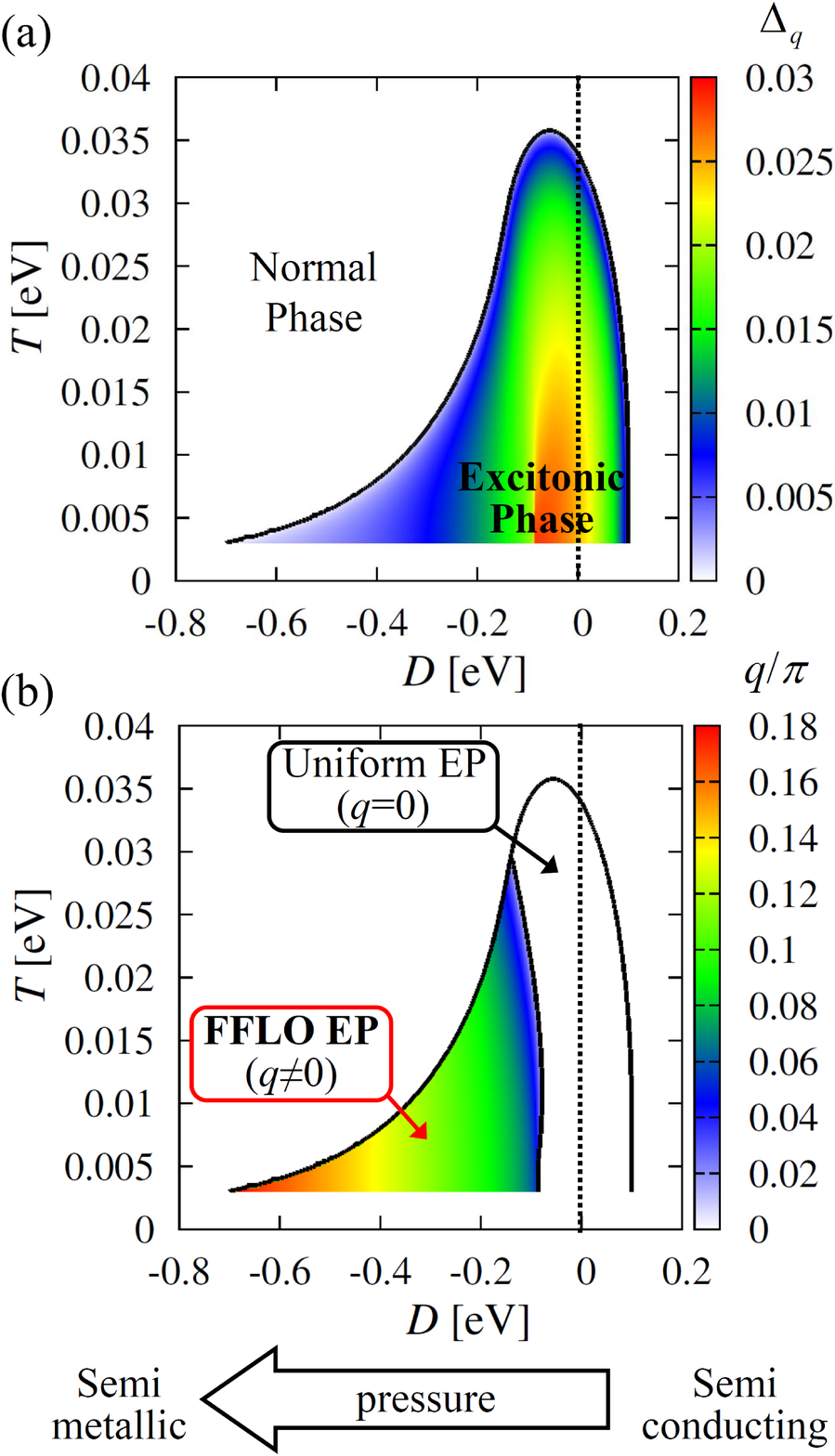}
\caption{(Color online) 
Excitonic phase diagrams of the three-chain Hubbard model for Ta$_2$NiSe$_5$ as functions of the energy gap $D$ 
and temperature $T$ for $n$=2 and $V$=0.4 eV, where the magnitudes of the order parameter $\Delta_{q}$ (a) and wavenumber $q/\pi$ (b) are shown. 
}
\label{Fig2}
\end{center}
\end{figure}

Figure \ref{Fig2}(a) shows the excitonic phase diagram on the $D-T$ plane, where the excitonic order with $\Delta_q \ne 0$ is realized for $D\siml 0.1$ below the transition temperature $T_c$. In the semiconducting case with a narrow gap between the $c$ and $f$ bands for $0<D\siml 0.1$, the transition from the semiconductor to the EI, i. e., the BEC of excitons, takes place as previously reported in Ref.\cite{PRB.87.035121}. When the gap $D$ decreases, $T_c$ rapidly increases with increasing carrier density, as expected in the BEC regime. $T_c$ still increases with decreasing $D$ in the semimetallic case with slightly overlapping $c$ and $f$ bands for $-0.06\siml D<0$, where the exciton binding energy $\sim \Delta_q$ is larger than the Fermi energy measured relative to the band edge $\sim |D|/2$. On the other hand, in the semimetallic case with relatively larger $c$-$f$ band overlapping for $D\siml -0.06$ where $\Delta_q$ is smaller than $|D|/2$, the transition from the semimetal to a BCS-like excitonic condensation takes place. In this case, $T_c$ gradually decreases with increasing the band overlapping $|D|$. Thus, the system shows a BCS-BEC crossover at $D\sim -0.06$, where $T_c$ shows a maximum as shown in Fig. \ref{Fig2}(a). 

In the semimetallic case with the different $c$-$f$ band degeneracy, where the band overlapping causes the imbalance of the Fermi wavenumber $k_{\rm F}^{c}\neq k_{\rm F}^{f}$, one can expect that the condensation of excitons with finite center-of-mass momentum $q$ takes place, analogous to that of Cooper pairs in FFLO superconductivity under an external magnetic field, where the Zeeman splitting causes the imbalance of the Fermi wavenumber for each spin $k_{\rm F}^{\uparrow}\neq k_{\rm F}^{\downarrow}$. In fact, the FFLO excitonic state with $q\ne 0$ is stabilized in a wide parameter region for the semimetallic case as shown in Fig. \ref{Fig2}(b), where the wavenumber $q$ for which the free energy $\delta F_{q}$ becomes minimum is plotted on the $D-T$ plane. The FFLO EP ($q\ne 0$) is observed for $D\siml -0.08$, while the uniform EP ($q=0$) is observed for $D\simg -0.14$, and the phase boundary between the two is located at $-0.14\siml D \siml -0.08$ depending on $T$.

\begin{figure}[t]
\begin{center}
\hspace*{-5mm}
\includegraphics[width=8.7cm]{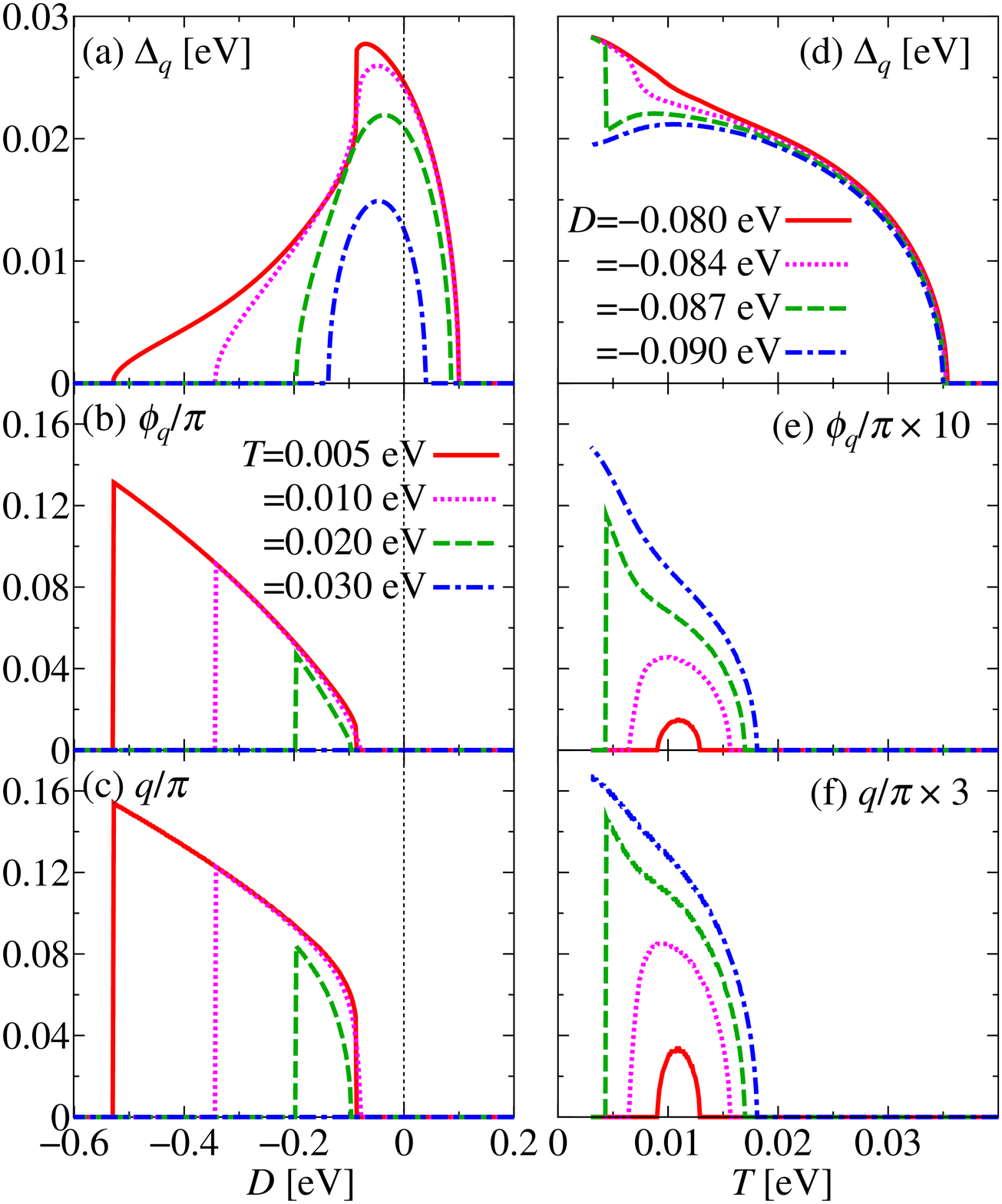}
\includegraphics[width=6.0cm]{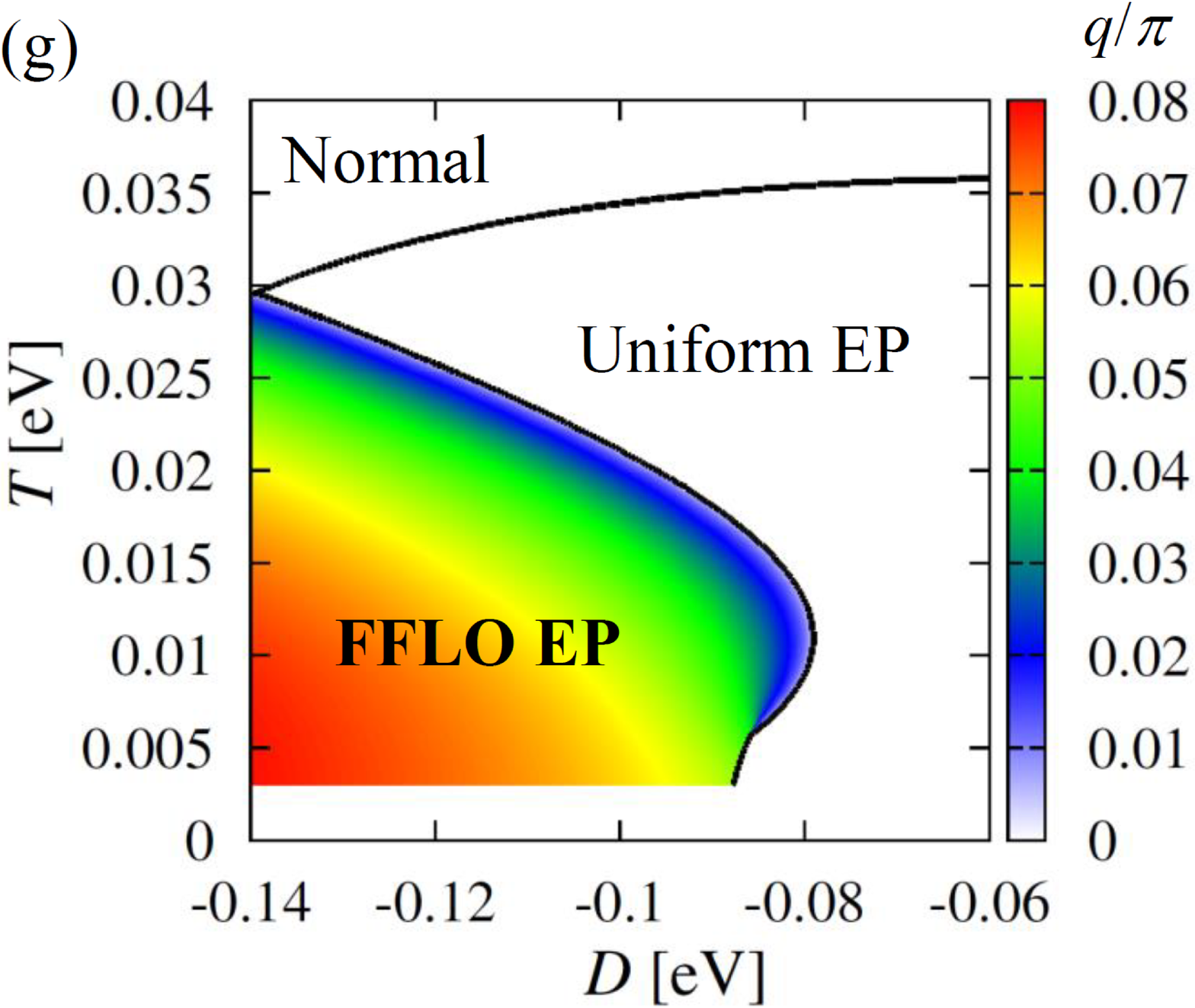}
\caption{(Color online) 
$D$-dependence of the magnitude of the excitonic order parameter $\Delta_{q}$ (a), its relative phase $\phi_q$ (b), and the wavenumber $q/\pi$ (c) for several values of $T$. $T$-dependence of $\Delta_{q}$ (d), $\phi_q$ (e), and $q/\pi$ (f) for several values of $D$. (g) Enlargement of Fig. \ref{Fig2}(b) around the phase boundary between the uniform and FFLO EPs. 
}
\label{Fig3}
\end{center}
\end{figure}

In Figs. \ref{Fig3}(a)-(c), we plot the magnitude of the excitonic order parameter $\Delta_{q}$, 
its relative phase $\phi_q$, and the wavenumber $q$ for which the free energy $\delta F_{q}$ becomes minimum as functions of $D$ for several values of $T$. In the EP, $\Delta_{q}$ becomes finite and increases (decreases) with decreasing $D$ in the BEC (BCS) regime and then shows a peak in the crossover region. In the FFLO EP, both $q$ and $\phi_q$ become finite and monotonically increase with decreasing $D$ towards the phase boundary with the normal phase. When approaching the transition from the EP to the normal phase, $\Delta_{q}$ continuously becomes zero, indicating a second-order phase transition. When approaching the transition from the FFLO EP to the uniform EP, both $\phi_q$ and $q$ continuously become zero at relatively high temperatures of $T=0.01-0.03$, where the transition is the second-order, while discontinuously become zero at $T= 0.005$ where $\Delta_{q}$ also shows a discontinuous jump indicating the first-order phase transition. Detailed calculations indicate that the transition between the uniform and FFLO EPs is second-order for $T\simg 0.006$ but first-order for $T\siml 0.006$, as shown in Fig. \ref{Fig3}(g), where a remarkable reentrant transition is observed at the phase boundary between the uniform and FFLO EPs as mentioned in detail below.

To observe the reentrant transition explicitly, we plot the $T$-dependence of $\Delta_{q}$, $\phi_q$, and $q$ for several values of $D$ in the narrow region of the uniform-FFLO phase boundary with $-0.09 \leq D\leq -0.08$ in Figs. \ref{Fig3}(d)-(f), respectively. 
For $D=-0.09$, both $\phi_q$ and $q$ monotonically increase with decreasing $T$ below the critical temperature $T=0.018$ at which the second-order phase transition between the uniform and FFLO EPs takes place. 
For $D=-0.08$ ($-0.084$), when $T$ decreases, we observe the reentrant transition 
at $T=0.013$ ($0.016$) from the uniform EP to the FFLO EP and 
at $T=0.009$ ($0.006$) from the FFLO EP to the uniform EP, 
where both transitions are found to be the second-order. 
On the other hand, for $D=-0.087$, when $T$ decreases, 
we observe the second-order phase transition from the uniform EP to the FFLO EP at $T=0.017$ 
but the first-order transition from the FFLO EP to the uniform EP at $T=0.004$, where $\Delta_{q}$ also shows a discontinuous jump. Around the first-order phase transition, we also confirmed that the free energy $\delta F_{q}$ has a double minimum with respect to $q$ (not shown). 

\begin{figure}[t]
\begin{center}
\includegraphics[width=6.0cm]{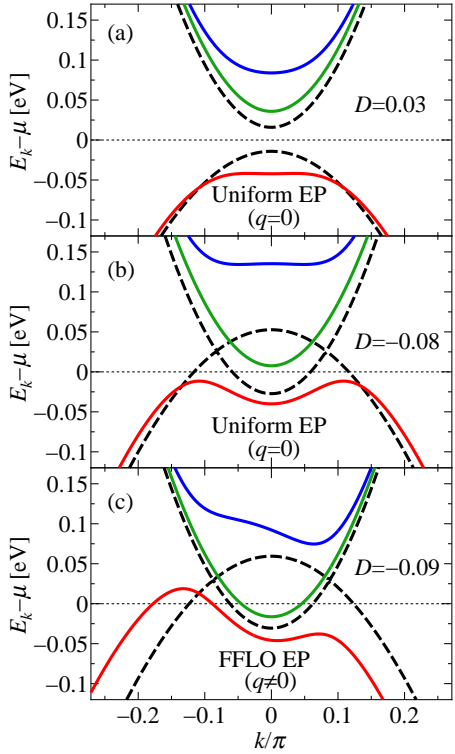}
\caption{(Color online) 
Energy band structures near the chemical potential $\mu$ as functions of wavenumber $k/\pi$ around the Brillouin zone center at $T=0.005$ eV: (a) EI in the uniform EP for $D=0.03$ eV, (b)  that for $D=-0.08$ eV and (c) excitonic semimetal in the FFLO EP for $D=-0.09$ eV. 
}
\label{Fig4}
\end{center}
\end{figure}

A significant difference between the uniform and FFLO excitonic states is the corresponding band dispersion given by Eq. (\ref{eq:EMF}), which yields a more explicit form upon using Eq. (\ref{eq:Delta}) as
\begin{equation}
E_{k,\pm}^{\rm MF}=\epsilon_{+}(k,q)\pm\sqrt{\epsilon_{-}^{2}(k,q)+4\Delta_{q}^{2}(1+{\rm cos}(k-\phi_{q}))}. 
\label{eq:EMF2}
\end{equation}
In Figs. \ref{Fig4}(a)-(c), the energy band structures near the chemical potential $\mu$ are plotted as functions of wavenumber $k/\pi$ around the Brillouin zone center at $T=0.005$ eV in the following three specific cases. Figure \ref{Fig4}(a) shows the energy band of the EI in the uniform EP with $\Delta_{q}=0.022$ and $\phi_q=q=0$ for the semiconducting case with $D=0.03$, where the flattening of the valence band top is observed, as shown in the previous theory\cite{PRB.87.035121}, which well accounts for the ARPES experiments on Ta$_2$NiSe$_5$\cite{PRL.103.026402,JSNM.25.1231}. For the EI in the 
semimetallic case with slight band overlapping 
with $D=-0.08$, where $\Delta_{q}=0.028$ and $\phi_q=q=0$ (uniform EP), the valence band top shows a double peak structure [see Fig. \ref{Fig4}(b)]. This is caused by the strong hybridization of $c$ and $f$ bands due to the excitonic condensation with large $\Delta_{q}$. In contrast, in the FFLO EP for $D=-0.09$ eV, where $\Delta_{q}=0.020$, $\phi_q/\pi=0.013$, and $q/\pi=0.053$, we observe the semimetallic band structure with a marked asymmetry with respect to $k=0$ [see Fig. \ref{Fig4}(c)]. This is caused by the hybridization of $c$ and $f$ bands with a wavenumber shift $q$ due to the imbalance of the Fermi wavenumber $k_{\rm F}^{f}-k_{\rm F}^{c}$ and also by the nontrivial wavenumber shift due to the relative phase of the order parameter $\phi_q$ as shown by Eq. (\ref{eq:EMF2}). Note that the transition between the EI in the uniform EP and the excitonic semimetal in the FFLO EP is first-order at the low temperature $T=0.005$ eV as shown in Figs. \ref{Fig3}(a)-(g).

In addition to the FFLO state with $q>0$ and $\phi_{q}>0$ mentioned above, another degenerate FFLO state exists with $-q$ and $\phi_{-q}=-\phi_{q}$, where the dispersion $E_{k,\pm}^{\rm MF}$ with $-q$ is equivalent to $E_{-k,\pm}^{\rm MF}$ with $q$ as can be seen from Eq. (\ref{eq:EMF2}). The two degenerate states are categorized into the Fulde-Ferrell (FF) type, in which the order parameter has a homogeneous magnitude but a modulated complex phase factor. This degeneracy may be resolved by various effects in real materials such as surface, impurity, and lattice distortion, resulting in the Larkin-Ovchinnikov (LO) type states in which the order parameter is real and spatially modulated. In fact, in the electron-hole bilayer system with density imbalance, the LO type state has been found to be stabilized in a finite-size system\cite{PRB.81.115329} compared with the FF type state, which was revealed by momentum space calculations\cite{PRB.75.113301,JPSJ.79.033001} similar to those in the present study. Therefore, discussing the possibility of the LO type excitonic states in the present model will be an interesting future problem.

Here, we briefly discuss the effect of the orthorhombic-to-monoclinic structural transition in Ta$_2$NiSe$_5$, 
which was found to be induced in the EI by taking account of the coupling $g$ between the electron and the uniform 
shear distortion $\delta$ of the chain\cite{PRB.87.035121}. Then, we consider the effect of the same electron-lattice 
coupling and obtain some preliminary results: the FFLO state is suppressed by $\delta$ as it resolves the conduction 
band degeneracy but survives up to a critical value $\delta_c$, for example, $g\delta_c \sim 0.01$ eV for $D=-0.1$ eV. 
Therefore, we expect that the FFLO state with small monoclinic distortion will be realized in semimetallic 
Ta$_2$NiSe$_5$ under high pressure, where the monoclinic phase is suppressed by pressure and finally disappears 
at a critical pressure\cite{pressure}. Detailed results with the explicit inclusion of the electron-lattice coupling 
will be reported in a subsequent paper.

In summary, we have investigated the three-chain Hubbard model for Ta$_2$NiSe$_5$ over the wide range of the energy gap $D$ between the twofold degenerate $c$ band and the nondegenerate $f$ band and have obtained the excitonic phase diagram on the $D-T$ plane, where the second-order phase transition from the normal phase to the excitonic phase occurs at $T_c$. There is a peak of $T_c$ in the crossover region between the BEC ($D\simg 0$) and BCS ($D\siml 0$) regimes. In the semimetallic case with $D<D_c<0$, where the band overlapping is larger than a critical value $|D_c|$, the imbalance of the $c$ and $f$ Fermi wavenumber due to the difference in the band degeneracy results in the remarkable FFLO excitonic state. This state is characterized by the condensation of excitons with finite center-of-mass momentum $q$ corresponding to the Fermi wavenumber imbalance. The band structure of the FFLO state is asymmetric with respect to $k=0$ owing to the wavenumber shift $q$ together with the relative phase of the order parameter $\phi_q$, in contrast to the uniform excitonic state with $q=\phi_q=0$ realized in the semiconducting ($D>0$) and slightly band overlapping semimetallic ($D_c<D<0$) cases. In these cases, flattening or a double peak structure of the valence band top is observed. With decreasing $D$, corresponding to increasing pressure, the system shows a first-order phase transition from the uniform state to the FFLO state at low temperatures 
while a second-order phase transition at relatively high temperatures. A reentrant uniform-FFLO-uniform transition is also observed as a function of $T$ for a fixed $D$ around $D_c$.

In the semiconducting case with $D>0$, our results regarding the EI are the same as the previous results 
in Ref.\cite{PRB.87.035121}, where the orthorhombic-to-monoclinic structural transition in Ta$_2$NiSe$_5$ at ambient pressure was well accounted for by the transition from the semiconductor to the EI, which shows the flattening of the valence band top as observed in the ARPES experiments below the transition\cite{PRL.103.026402,JSNM.25.1231}. 
The present results for the semimetallic case with $D<0$ including the FFLO excitonic state have been obtained from a straightforward extension of the semiconducting case and are therefore expected to be realized in Ta$_2$NiSe$_5$ under high pressure as $D$ is considered to be a decreasing function of pressure. 
In fact, Ta$_2$NiSe$_5$ becomes semimetallic under high pressure and also shows the orthorhombic-to-monoclinic structural transition, which is suppressed by pressure and finally disappears at the critical pressure around which superconductivity is observed\cite{pressure}. 
Our preliminary calculation with the random phase approximation revealed that the superconductivity occurs due to the enhanced excitonic fluctuation towards the excitonic phase boundary. 
Explicit results for the superconductivity as well as detailed results for the FFLO excitonic state including thermodynamic, transport, and optical properties will be reported in subsequent papers.

\begin{acknowledgments}
One of us (Y. \=O) would like to express his sincere thanks to Professor H. Fukuyama for directing his attention to the present problem and for helpful suggestions. We would also like to thank Y. Ohta, T. Kaneko, K. Sugimoto, and J. Ishizuka for valuable comments and discussions. This work was partially supported by a Grant-in-Aid for Scientific Research from the Ministry of Education, Culture, Sports, Science and Technology. 
\end{acknowledgments}


\bibliography{17165}

\providecommand{\noopsort}[1]{}\providecommand{\singleletter}[1]{#1}%
\begin{thebibliography}{10}

\bibitem{Knox.1963}
R.~Knox: in{\em ``Solid State Physics"}, ed. F.~Seitz and D.~Turnbull (Academic
  Press, New York, 1963), Suppl. 5, p. 1.

\bibitem{PR.158.462}
D.~J$\acute{\rm e}$rome, T.~M. Rice, and W.~Kohn: Phys. Rev {\bfseries 158}
  (1967) 462.

\bibitem{RMP.40.755}
B.~I. Halperin and T.~M. Rice: Rev. Mod. Phys {\bfseries 40} (1968) 755.

\bibitem{Inorg.Chem.24.3611}
S.~A. Sunshine and J.~A. Ibers: Inorg. Chem. {\bfseries 24} (1985) 3611.

\bibitem{JLCM.116.51}
F.~J. DiSalvo, C.~H. Chen, R.~M. Fleming, J.~V. Waszczak, R.~G. Dunn, S.~A.
  Sunshine, and J.~A. Ibers: J. Less-Common Met. {\bfseries 116} (1986) 51.

\bibitem{PRL.103.026402}
Y.~Wakisaka, T.~Sudayama, K.~Takubo, T.~Mizokawa, M.~Arita, H.~Namatame,
  M.~Taniguchi, N.~Katayama, M.~Nohara, and H.~Takagi: Phys. Rev. Lett
  {\bfseries 103} (2009) 026402.

\bibitem{JSNM.25.1231}
Y.~Wakisaka, T.~Sudayama, K.~Takubo, T.~Mizokawa, N.~L. Saini, M.~Arita,
  H.~Namatame, M.~Taniguchi, N.~Katayama, M.~Nohara, and H.~Takagi: J.
  Supercond. Nov. Magn. {\bfseries 25} (2012) 1231.

\bibitem{PRB.87.035121}
T.~Kaneko, T.~Toriyama, T.~Konishi, and Y.~Ohta: Phys. Rev. B {\bfseries 87}
  (2013) 035121.

\bibitem{PRB.93.041105}
K.~Sugimoto, T.~Kaneko, and Y.~Ohta: Phys. Rev. B {\bfseries 93} (2016)
  041105(R).

\bibitem{PRB.90.155116}
K.~Seki, Y.~Wakisaka, T.~Kaneko, T.~Toriyama, T.~Konishi, T.~Sudayama, N.~L.
  Saini, M.~Arita, H.~Namatame, M.~Taniguchi, N.~Katayama, M.~Nohara,
  H.~Takagi, T.~Mizokawa, and Y.~Ohta: Phys. Rev. B {\bfseries 90} (2014)
  155116.

\bibitem{PRB.75.113301}
P.~Pieri, D.~Neilson, and G.~Strinati: Phys. Rev. B {\bfseries 75} (2007)
  113301.

\bibitem{JPSJ.79.033001}
K.~Yamashita, K.~Asano, and T.~Ohashi: J. Phys. Soc. Jpn {\bfseries 79} (2009)
  033001.

\bibitem{PRB.81.115329}
J.-X. Zhu and R.~Bishop: Phys. Rev. B {\bfseries 81} (2010) 115329.

\bibitem{pressure}
K.~Matsubayashi: private communication .

\end{thebibliography}
\end{document}